\documentstyle[epsf]{article}
\title{On Oscillations in Cellular Automata}
\author{Jan Hemmingsson$^{a,b}$ and Hans J.~Herrmann$^{a}$ \\ $^{a}$ HLRZ,
	Forschungzentrum J\"ulich\\ Postfach 1913, W--5170 J\"ulich, Germany\\
$^{b}$IFM, 		Link\"oping Institute of Technology\\ S--58183 Link\"oping,
Sweden}
\date{\today}
\maketitle
\begin{abstract}
\noindent
We investigate cellular automata where some global quantity varies
periodically or quasiperiodically
with time. We find that these systems are highly predictable,
and can be rather well described by a set of mean-field variables.
We conclude that this is not a collective phenomenon --- where different
subsystems are
supposed to synchronize --- but rather like many very weakly
coupled oscillators fluctuating around one exact frequency. The global quantity
measured is a mean taken over all these subsystems, and gets more precise
the larger the system is.
\end{abstract}
\begin{document}
\noindent
\raisebox{11.5cm}[][]{preprint HLRZ 99/92}
\\
\noindent
\flushbottom
\\
\noindent
The temporal behaviour of spatially extended systems of many degrees
of freedom has recently attracted much interest. One question of
special importance is under which circumstances a global quantity can
show a non-transient evolution. It has been argued \cite{grinstein} that
isotropic systems governed by local interactions only, cannot exhibit
non-trivial global behaviour. Despite these arguments, there are
examples of cellular automata (CA) with periodic or
quasiperiodic oscillations \cite{chate,jan}. Let us start by
reviewing the main features of these systems.
\\
\\
On a regular lattice of size $L^{d}$, an isotropic, deterministic,
two-state CA will be
defined by a totalistic rule, i.e. the state (0 or 1)
of each cell is determined by the sum of all spins
in the nearest neighbourhood, the site itself included.
All cells are updated synchronously, and we normally use periodic
boundaries, although it does not affect the results.
We consider here `rule 33' in three dimensions \cite{jan}, which can
be described as follows:
\begin{equation} \label{min_automat}
\sigma_{i}(t+1)= \left\{
\begin{array}{ll}
1 & \mbox{if $h_{i}(t)=0$ or $h_{i}(t)=5$} \\
0 & \mbox{otherwise.}
\end{array}
\right.
\end{equation}
The global quantity of main interest is the fraction of sites with
spin 1, the `magnetization'. Normaly, the initial configuration is set
up by randomly assigning 0 or 1 to each cell with a certain
probability, often chosen to be 0.5. The magnetization at time $t+1$ is
plotted against the magnetization at time $t$, a so called return map.
The result is a noisy limit cycle, which gets less noisy as the system
size grows. The limit cycle does not seem to shrink with increasing
system size, and there are no experimental evidences this to be a transient.
For the limit cycle, there exists a winding number $w=-0.3371$, defined
as the number of
revolutions made per time step, the minus sign indicating that
consecutive plots will describe a clock-wise motion in the phase
space. The intrinsic noise was found to be stochastic, in the sense
that the nearest neighbour distance of points in the limit cycle goes
like $M^{-1/2}$ where $M$ is the number of points
in the plot. The system is also very stable against
external noise and changes in the initial configuration.
\\
\\
How does the temporal behaviour of the magnetization depend on the
initial configuration? Time-series analysis has been used for these
systems to characterize
the periodicity \cite{binder}, but until now, no systematic investigation of
the transient has been performed. We simulated the automaton of
eq.\ref{min_automat} for sizes up to $512^{3}$ on a Connection Machine CM-2.
Let us introduce an angle
$\phi(t)=\arctan ( \frac{m(t+1)-0.3}{m(t)-0.3})$ and measure its
distribution for different initial configurations of the same
magnetization. In figure \ref{phi(500)} the distribution of
$\phi(500)$ for different system sizes is plotted. Note that the $x$-axis
is very stretched. Even for the smallest systems in the figure,
$32^{3}$, the
distribution is accurate. The larger the system is,
the more predictable is the evolution of the magnetization.
Thus it seems that for an infinite system there is exactly the same
time sequence for the magnetization for each initial configuration with
given initial magnetization and correlations. The underlying mechanism
seems to be local, and interactions between subsystems less
important. That would also explain why the intrinsic noise is
stochastic, since it is then just the result of summing
the magnetization over all the subsystems. All this indicates that the
quasiperiodicity is intrinsic to a small system, characterized by just
a few variables, and not due to some collective synchronization of the
entire system.
\begin{figure}[htb]
\centerline{
\epsfxsize=6.3cm
\epsfbox{phi500.eps}
}
\caption{The distribution of the angle $\phi(t)$ after 500 time steps
for different lattice sizes $L$. The $x$-axis is in radians.
\label{phi(500)}}
\end{figure}
\\
\\
Which are the relevant variables of the system? We have already
mentioned the magnetization. The structures of zeros and ones that can
be seen in a snapshot of
the system \cite{jan} lead us to introduce a set of point
correlation functions $c_{k}$ by summing zeros surrounding a spin zero
and ones
surrounding a spin one. Let $N$ be the number of sites in the system.
For each correlation function, all the $S(k)$ sites that are $k$th
nearest neighbours of the central site are taken into account:
\begin{equation}
c_{k} = \frac{1}{N S(k)} \sum_{i}^{N} \sum_{n}^{ S(k)}
\delta_{\sigma_{i} \sigma_{n}}
\end{equation}
These neighbouring sites form a shell surrounding the
central site.
For a better visualization, the correlation functions and the
magnetization
are plotted for every third time step in figure \ref{correlation}a. It
can
be seen that the correlation functions have the same quasiperiodic
behaviour as the magnetization, but with a certain phase shift. The
figure also
shows that the amplitude of the correlations decreases with
increasing order. It therefore seems quite natural that these
correlations characterize the (random) nature of the configuration of
a subsystem that at the end determine the entire quasiperiodic
sequence of its evolution.
\begin{figure}[htb]
\centerline{
   	\epsfxsize=6.3cm
	\put(7, 100){(a)}
	\epsfbox{correlation.eps}
	\hfil
	\epsfxsize=6.3cm
	\put(7, 100){(b)}
	\epsfbox{meanfield.eps}
	}
\caption{ (a): Plotted are the magnetization (top) and the
correlations
$c_{1}$ to $c_{4}$ (one below the other) as functions of time, showing
only every third time-step measured from numerical simulations of
the CA with $L=128^{3}$. (b): The inner solid line is the mean field
iteration including the nearest neighbour correlation
(eq.~3), the outer line also includes next nearest
neighbour correlation. The points come from a direct simulation of the
CA for $L=256^{3}$. \label{correlation} }
\end{figure}
\\
It is of interest to try to describe the system with a mean field
approach, using these variables in order to calculate the
global properties one
measures in the real system. Several attempts have already been made
\cite{binder_again,jason}, but none very successful. Encouraged by the
evidences of a deterministic evolution of the magnetization and the
correlations,
we make an ansatz that the magnetization at
time $t+1$ can be described as a polynomial $P_{0}(m(t), c_{1}(t),
c_{1}(t),
...)$ of all permutations of the magnetization and the correlation
functions at time $t$. Equivalently, there will be polynomials
$P_{k}(m(t),
c_{1}(t), c_{2}(t), ...)$ as expressions for the correlations.
These polynomials do not explicitly depend on the
time $t$, but only tell us how to go from $t$ to $t+1$. For each
$P_{k}$,
we have to solve an overdetermined system of equations, and we use the
least
square method to find the coefficients.
The result can be written as a set of iterative
functions. An example for the resulting equations for the
magnetization and the first
correlation function:
\begin{equation}
\begin{array}{lll} \label{mf_equation}
m(t+1)=
& 0.718+0.578 c_{1}(t) - 0.580 c_{1}^{2}(t) - 3.016 m(t)
\\ & +1.883 m(t)
c_{1}(t)+2.891 m^{2}(t)
\\
c_{1}(t)=
& 0.660-1.660 c_{1}(t)+1.499c_{1}^{2}(t)-1.421 m(t) \\ &+3.472 m(t)
c_{1}(t)-0.188 m^{2}(t)
\end{array}
\end{equation}
The coefficients have been rounded to three decimals.
\\
\\
As a comparison to the real system, we have  plotted in figure
\ref{correlation}b the real limit cycle (points) together with mean
field iterations
taking one (the inner line) and two (the outer line) correlation
functions into account.
Both mean field iterations are quasiperiodic with
a period close to three. It is striking how well the limit cycle in
figure \ref{correlation}b is described by just three variables. This
very
much corroborates our picture.
\\
\\
Let us finally return to the original question. The reason why
these systems violate the statements in \cite{grinstein} is rather
simple:
The global behaviour seen here is not  a collective phenomenon. We
have seen that any system larger than a given size --- about 30
lattice spacings --- has
a well defined trajectory in a reduced `phase space', in which each
configuration is random in its detail, but characterized by a finite
set of correlation functions. Each subsystem of a larger
system oscillates `on its own', with fluctuations around an exact
frequency as a result of small statistical fluctuations in the value
of the
correlations, due to their random nature.
An average taken over all the subsystems will thus be more precise the
larger
the system is. Since there are no well defined collective states of
the system, the arguments
of \cite{grinstein} do not apply.
The system is simply composed of essentially uncoupled local
oscillators, so coupled map lattices do not seem to bee the adequate
description either. The picture described here seems to be the key for
understanding the rather puzzling phenomenon of global oscillations
found not only in high dimensional automata, but also in many natural
phenomena like the Belousov-Zhabotsinski reaction.
\\
\\
We thank our colleagues at HLRZ and at the Scientific Visualization
Department of HLRZ at GMD, Bonn for discussions. Special thanks for
special help to Peter Ossadnik.
This work was supported in part by the Swedish Natural Research
Council and
the PHC foundation, Brunflo.
\\

\end{document}